\documentclass[12pt]{article}
\usepackage{graphicx}
\usepackage{epsfig}
\usepackage{epstopdf}
\usepackage{amsmath}
\usepackage{amsfonts}
\usepackage{amssymb}
\usepackage{ccaption}
\usepackage[usenames]{color}

\usepackage[
      colorlinks=true,
      linkcolor=blue,  
      citecolor=blue,
      urlcolor=blue,    
      filecolor=blue,     
      linkcolor=blue,
      pdfstartview=FitV,
      pdftitle={},
        pdfauthor={Hubeny},
        pdfsubject={bulk probes},
        pdfkeywords={AdS/CFT},
        pdfpagemode=None,
        bookmarksopen=true    
      ]{hyperref}

\definecolor{labelkey}{gray}{0.1}
  \captionnamefont{\bfseries}
  \captiontitlefont{\small\sffamily}
  \captiondelim{: }
  \hangcaption

\def\fig#1{Fig.\,\ref{#1}}
\def\req#1{(\ref{#1})}

\def\Om{\Omega}

\def\({\left(}
\def\){\right)}

\def\CH{{\cal H}}

\def\CN{{\cal N}}
\def\CO{{\cal O}}

\def\CS{{\cal S}}

\def\RR{\mathbb{R}}

\def\rh{r_+}

\def\SAdS#1{Schwarzschild-AdS$_{#1}$}

\def\Sp{{\bf S}}

\def\ord#1{\CO\!\left( #1 \right)}
\def\abs#1{\mid \! #1 \! \mid}
\def\rh{r_+}
\def\zh{z_+}

\def\scri{\mathcal I}

\numberwithin{equation}{section}

\begin{document}

\setlength{\unitlength}{1mm}

\begin{titlepage}

\begin{flushright}
DCPT-11/07
\end{flushright}
\vspace{1cm}

\begin{center}
{\bf \Large Holographic insights and puzzles}
\end{center}

\vspace{4mm}
\begin{center}
Veronika E.~Hubeny$^{a}$

\vspace{.4cm}
{\small \textit{$^{a}$Centre for Particle Theory \& Department of Mathematical Sciences, Durham University, }}\\
{\small \textit{Science Laboratories, South Road, Durham DH1 3LE, United Kingdom}} \\
\vspace*{0.4cm}
{\small {\tt veronika.hubeny@durham.ac.uk}}
\end{center}

\vspace{5mm}

\begin{abstract}

The talk is composed of two parts, both set within the AdS/CFT context.  In the first part, I discuss holographic insight into strongly coupled field theory in a black hole background.  I conjecture two new gravitational solutions, dubbed black funnels and black droplets, which describe two distinct deconfined phases in the field theory at finite temperature.  I also briefly mention puzzles associated with an analogous set-up in a rotating black hole background.
In the second part of the talk, I discuss time-dependent states in a CFT on flat spacetime background, exemplified by the conformal soliton flow.  Here I focus on puzzles regarding the nature of entropy in time-evolving states and its holographic dual.

\end{abstract}

\end{titlepage}

\newpage

\section{Introduction and motivation}
\label{s:intro}


This talk is set within the context of the AdS/CFT duality, hence the word {\it holographic} in the title.  It concerns several problems which I have been recently intrigued by, and I wish to share with you some of the insights which I've gained from working on them.  But in order to make this talk a bit more interesting, I will also try to make it a bit more provocative: I will emphasize some of the puzzles, hitherto unresolved, which we have encountered  along the road.

To give you a broader perspective, after motivating the context, I will discuss two separate problems, both related to behaviour of strongly-coupled field theories and their gravitational duals in certain unconventional contexts.  In the first part, based on \cite{Hubeny:2009ru}, 
I will focus on strongly coupled field theory in a black hole background.  I will conjecture two new gravitational solutions, dubbed black funnels and black droplets, which describe two distinct deconfined phases in the field theory at finite temperature.  I will then briefly mention puzzles associated with an analogous set-up in a rotating black hole background.
In the second part of the talk, based on \cite{Figueras:2009iu}, I will discuss time-dependent states in a CFT on flat spacetime background, exemplified by the conformal soliton flow.  Here I will focus on puzzles regarding the nature of entropy in such time-evolving state and its holographic dual.

\paragraph{Motivation from a field theory perspective:}
Over the last few decades we have learned that quantum field theory on curved spacetime exhibits a large variety of fascinating effects, such as vacuum polarization, particle production, etc.\footnote{
See e.g.\ \cite{Birrell:1982ix,Jacobson:2003vx,Ross:2005sc}  for reviews. 
}
 In particular, the discovery of Hawking radiation by black holes has led to deep and fundamental issues, which have played an important role in attempts to formulate a quantum theory of gravity. However, much of our understanding is anchored in the context of free field theories. The details, and even the relevant degrees of freedom or the basic physical processes, are much less clear when the field theory interacts strongly.  Clearly, further exploration of this territory carries promise of exciting new revelations.  An especially interesting background to consider is a black hole spacetime. 
Our first broad goal, then, is to understand strongly coupled field theory on black hole backgrounds.

Let us first recall  the situation for {\it free} fields on a fixed black hole spacetime.    As is well known,
the black hole radiates and absorbs field theory quanta in a manner much like an ideal black body at the Hawking temperature $T_H$.  
The best-understood state of the field theory is the Hartle-Hawking vacuum, which describes radiation in equilibrium with the black hole, attained by balancing the ingoing and outgoing fluxes.
The renormalized stress tensor in this state is regular on both future and past horizons, and the state is conveniently defined by a Euclidean path integral which is periodic in imaginary time. 
While the Hartle-Hawking state is well-defined only for static (but not more general stationary) black holes, 
one can also consider the time-asymmetric Unruh vacuum which describes only the outgoing flux of Hawking quanta from the black hole.  In such settings, one can compute occupation numbers and the stress tensor far from the black hole, though this becomes more difficult closer to the horizon.  Increasing the coupling, one would expect this basic picture to remain valid when the field theory quanta interact weakly, at least in perturbation theory.

In contrast,  for {\it strongly interacting} field theories on black hole backgrounds, it is a challenge to obtain even a qualitative understanding of Hawking radiation.   One can still formally define the Hartle-Hawking state via a Euclidean path integral and, due to periodicity in Euclidean time, it will satisfy the KMS condition, so that in this sense the state is thermal.
In addition, far from the black hole this construction reduces to the thermal path integral in flat space, giving a manifestly thermal
stress tensor there.  However, more detailed information is sorely lacking, and even less is understood regarding the corresponding Unruh states or other out-of-equilibrium situations.\footnote{
See e.g., \cite{Carter:1976di,FH,MacGibbon:1990zk,MacGibbon:1991tj} for further comments.}

\paragraph{AdS/CFT to the rescue:}
As became increasingly more apparent during the last decade, the Anti-de Sitter/conformal field theory (AdS/CFT) duality \cite{Maldacena:1997re,Gubser:1998bc,Witten:1998qj} offers an invaluable tool to probe field theories at strong coupling.
This correspondence maps the dynamics of the field theory to string theory (or classical gravity) in a higher dimensional Anti-de Sitter (AdS) spacetime. The field theories in question are typically non-abelian gauge theories; denoting the rank of the gauge group by $N$ and the 't Hooft coupling by $\lambda$, in the limit of large rank and strong 't Hooft coupling  ($N, \lambda \gg 1$)  the dual description of the field theory reduces to classical gravitational dynamics in the higher dimensional AdS  spacetime.\footnote{
For example, in the case of ${\cal N} = 4$ super Yang-Mills dual to AdS${}_5 \times \Sp^5$, the 't Hooft coupling $\lambda$ controls the curvature radius of the AdS spacetime in string units, while the string interactions are governed by $g_s \sim \frac{1}{N}$. Classical gravity is a good description when string interactions are suppressed and the curvatures are small. 
} 
Since the gravity and the field theory live in different number of dimensions, we refer to such duality as holographic; following standard terminology, we refer to gravity as living in the {\it bulk} of AdS and the field theory as living on the {\it boundary} of AdS.

A wonderful aspect of the AdS/CFT duality is that we can use one side of the correspondence to learn about the other; and this information-flow proceeds in both directions:
We can use classical gravity to gain much needed insight about behaviour of strongly coupled field theory, such as its transport properties and its phase structure.
Conversely, the field theory, albeit strongly coupled, provides a definition of quantum gravity with asymptotically AdS boundary conditions.  Hence even in extreme regimes where classical gravity breaks down, and  where we don't yet have the tools to understand the physics within string theory directly, the field theory remains well-defined.
This means that once we achieve sufficient understanding of the dictionary between the two sides, we can elucidate the long-standing quantum gravitational puzzles by re-casting them into the field theoretic language.  

\paragraph{Motivation from a quantum gravity perspective:}
Let us briefly indicate some of the gravitational questions we ultimately hope to answer within this programme, though they are not the main focus of this talk.  A central question of quantum gravity concerns the fundamental nature of spacetime. We have come to realize that spacetime is an emergent concept, but exactly how it emerges remains a mystery.
For instance, we would like to understand such basic notions as:  Which CFT configurations admit a dual description in terms of classical spacetime?  What types of spacetime singularities are physically allowed?  How are the disallowed singularities resolved?  How is spacetime causal structure encoded in the dual field theory?
Emergence of time is, if anything, even more mysterious than the emergence of space; indeed, conceptually the problem of time is one of the deepest problems in quantum gravity.
Even at the classical level, the non-uniqueness of bulk foliation corresponding to a given boundary foliation makes apparent the lack of any straight-forward mapping.
Nevertheless, even though the map between the boundary time and the bulk time is highly mysterious and probably quite non-local, 
we still expect that time-dependence on the boundary will be manifested by time-dependence in the bulk, and vice-versa.  Indeed, focusing on certain characteristic features of a given time-dependence can enable us to elucidate the gauge/gravity dictionary further. 
The second part of the talk will accordingly explore two main themes: Fluid/Gravity correspondence far from (local) equilibrium, and the holographic description of entropy in dynamical setting.

\paragraph{The fluid/gravity correspondence and beyond:}
In the large $N, \lambda$ regime of interest, we can further simplify the gauge/gravity description by focusing on the effective long-wavelength regime.  On the field theory side, this is given by fluid dynamics.  By definition (namely due to the very fact that we can meaningfully describe the system in terms of fluid variables), the boundary fluid is in local thermodynamic equilibrium -- nevertheless, the fact that it need not be in global equilibrium is the crucial ingredient which allows us to discuss time dependence.  
From the bulk standpoint, finite temperature of the boundary fluid maps to the existence of a corresponding bulk black hole horizon.  
Hence the long-wavelength dynamics of an interacting quantum field theory is given by gravitational dynamics of an asymptotically AdS black hole.  
This correspondence, explicitly constructed in 
\cite{Bhattacharyya:2008jc}, has become known as the {\em
  fluid/gravity correspondence}. The fluid/gravity correspondence
generalizes the extensive discussion of hydrodynamics of field
theories with gravitational duals\footnote{
For a review and references to earlier works on hydrodynamic aspects of $\CN =4$ Super Yang-Mills see  \cite{Son:2007vk}. Extensions of the fluid-gravity correspondence to include forcing, charge transport, etc., have been considered in many subsequent works; see e.g.\ \cite{Rangamani:2009xk,Hubeny:2010ry,Hubeny:2010wp} for recent reviews.
 }
and explicitly constructs spacetime
geometries dual to fluid flows in the hydrodynamic (i.e.\ locally equilibrated) regime. This provides a new approach to the calculation of properties such as transport coefficients of the field theory fluid.

Nevertheless, there are many situations of interest which do not belong to the long-wavelength regime.  For example, one can ask, {\it What about rapidly-varying backgrounds?}
In particular, consider the field theory on a Schwarzschild black hole background.  Far away, the equilibrium state is simply a thermal state at a temperature given by the black hole's temperature, which in turn determines the microscopic scale $\ell_{\rm mfp} \sim 1/T_{\rm BH}$.  But the curvature of the background geometry is of that same scale near the black hole.  
From the dynamical viewpoint, the fluid near the black hole wants to fall in on the timescale also  set by the black hole, i.e.\ $\ell_{\rm mfp}$.  We will see that considering such system in equilibrium with the outgoing Hawking quanta leads to an interesting bulk dual: as I will review momentarily, \cite{Hubeny:2009ru} suggested that the dual geometry exhibits new (yet to be found) solutions, dubbed black funnels and black droplets.\footnote{
As a follow-up, evidence for lower-dimensional realization of these was analyzed in \cite{Hubeny:2009kz} and further discussion of field theory on AdS black hole background was presented in \cite{Hubeny:2009rc}.
In fact, similar tools were recently exploited in \cite{Marolf:2010tg} to study field theory on de Sitter background.
} 

Another important type of situation which requires going beyond the fluid/gravity regime is one with sufficiently strong time-dependence. 
In general, the flow of a viscous fluid, which involves dissipation, necessarily leads to entropy production.    In the geometric description of the fluid flow, it was shown in \cite{Bhattacharyya:2008xc} that the event horizon in the spacetimes dual to non-linear fluid flows  provides a simple geometric construction for a 
Boltzmann H-function in the bulk geometry.  More specifically, the area-form of this event horizon when pulled back to the boundary was shown to lead to a natural local entropy current with non-negative divergence as required by the second law.
Rather intriguingly, the location of the event horizon in the bulk can be determined essentially locally despite the teleological nature of event horizons.  In \cite{Bhattacharyya:2008xc}, this was achieved
by assuming slow temporal variations and that the geometry will settle down to a stationary configuration at late times (which is of course natural as one expects the dissipation to cause the fluid  to achieve global equilibrium).  

In the second part of the talk, I wish to explore the question, {\it What if the geometry does not settle down?}
In particular, I will review the work of \cite{Figueras:2009iu}, which determines the location of the horizons in certain time-dependent geometries that do not
settle down to stationary finite-temperature solutions at late
times. Although our interest in this question originated from the
Bjorken flow geometry, here I will focus on the conformal
soliton geometry, which provides a simpler example with stronger
time dependence.

\section{Field theory on static black hole background}
\label{s:funnels}

Let us start by considering a thermal state of a strongly coupled CFT on a curved background.  Before delving into discussing a black hole background, I will briefly review the more familiar setup involving just a conformally flat background.

In particular let us consider global AdS, for which the boundary metric is spatially compact, since in such a case there is a sharp phase transition, the confinement/deconfinement transition,  in the planar (large $N$) limit. The CFT then lives on the Einstein Static Universe, $S^{3}\times \RR^1$.  There are two classes of static bulk geometries with such asymptotics:  the thermal AdS (which is metrically pure AdS) and the \SAdS{} black hole.
\begin{figure}
\begin{center}
\includegraphics[width=5in]{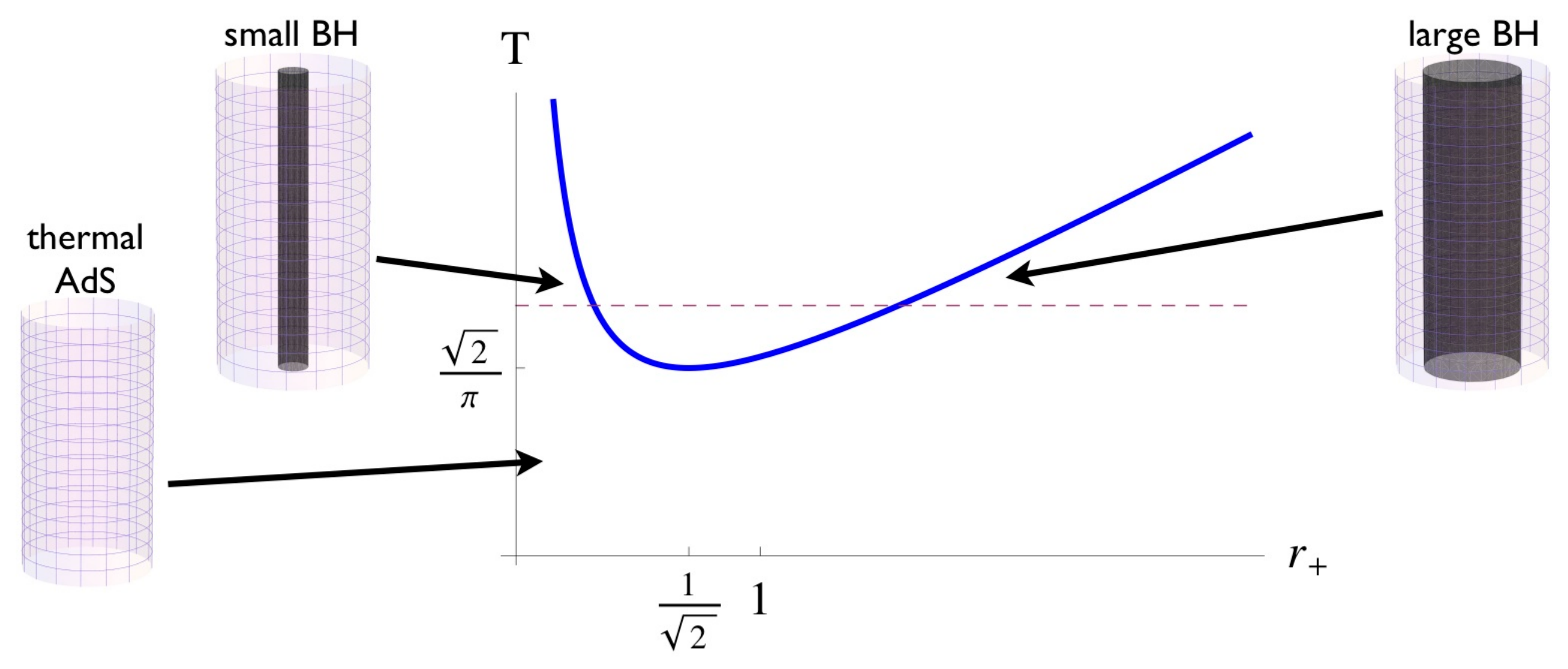}
\caption{Hawking-Page transition.  At small temperatures, where free energy is $\ord{1}$ and the field theory is in a confined phase, the only allowed static geometry with the requisite boundary conditions is thermal AdS, as indicated by the lower-left cylinder, showing one spatial (angular) direction and the time direction of the AdS boundary.  Above the deconfinement temperature, where the free energy is $\ord{N^2}$, two further geometries are possible, describing small and large black holes (with horizons indicated in the corresponding plots).  The large black hole dominates, while the small black hole is thermodynamically unstable.}
\label{f:HawkPage}
\end{center}
\end{figure}
\fig{f:HawkPage} shows the black hole temperature $T$ as a function of its horizon size $r_+$.  Unlike its asymptotically flat cousin, the \SAdS{} black hole has a minimal temperature given by the AdS scale, which occurs when the black hole size nears the AdS scale.  While smaller black holes are thermodynamically unstable, the large ones have positive specific heat.
As one increases the temperature, the thermal AdS$_{5}$ geometry exchanges dominance with the \SAdS{5} spacetime describing the large black hole \cite{Witten:1998zw}.
This transition is known as the Hawking-Page transition \cite{Hawking:1982dh}.
From the field theory viewpoint, the Hawking-Page transition corresponds to the confinement-deconfinement phase transition:
at low temperature we have the confined phase with $\ord{1}$ free energy and at high temperatures a deconfined phase with $\ord{N^2}$ free energy.\footnote{
A similar confinement-deconfinement transition can be seen at weak coupling as well  for four dimensional large-$N$ gauge theories on compact spatial manifolds \cite{Sundborg:1999ue,Aharony:2003sx}. 
}
The phase transition occurs at the deconfinement temperature $T_D$, set by the curvature radius of the $S^3$ (which is the only other dimensionful parameter in the problem if we restrict attention to conformal field theories).

\paragraph{AdS/CFT with curved boundary:}
We now wish to find the analog of this story for the case where the field theory does not live on the conformally flat  Einstein Static Universe $S^{3}\times \RR^1$, but rather a more non-trivially curved background, for instance one containing a black hole.
Here too, we can use the AdS/CFT correspondence, generalized to asymptotically-locally-AdS bulk geometries.
The previous literature on this subject has largely focussed on the case where the large-$N$ field theory is coupled to dynamical gravity.  This situation is holographically dual \cite{Gubser:1999vj} to Randall-Sundrum braneworld models  \cite{Randall:1999vf}, where much of the interest centered on possible implications for brane-world phenomenology. 

It will however be more useful for the present considerations to focus on the limit in which gravity on the brane decouples, i.e.\ the field theory propagates on a fixed non-dynamical black hole background.  From the bulk perspective, this is the limit where the brane recedes to the AdS boundary  and the metric on the brane becomes simply a boundary condition that is imposed by hand.   
In such cases, we are free to consider arbitrary black hole geometries which need not satisfy any equations of motion.   For example, we may consider a black hole background wherein we specify the horizon size $r_+$ and its temperature $T$ independently.   In terms of more familiar examples, one think of this analogously to Reissner-Nordstrom  black holes where one may adjust the relative size of $r_+$ and $T$ by varying the charge.
We refer to these arbitrarily specified black holes as {\em boundary black holes}  to distinguish them from {\em bulk} black holes that arise in the dual AdS description; the latter will satisfy Einstein's equation with a cosmological constant as well as a boundary condition determined by the field theory black hole metric.

As we have seen in the previous example with field theory formulated on $S^{3}\times \RR^1$, there may be multiple bulk solutions satisfying the specified boundary conditions.  These are distinguished by the induced boundary stress tensor, which corresponds to different phases of the field theory.
We will see that in order to unravel the phase structure, it will be particularly useful to consider $T \, r_+$ as a free parameter.
Nevertheless, to motivate our expectations,  we start by  considering our
 field theory as living on the standard 4-dimensional Schwarzschild geometry, which is in any case most naturally relevant for e.g.\ astrophysical considerations.

The only explicitly-known asymptotically locally AdS solutions with Schwarzschild boundary black hole is the AdS black string solution of \cite{Chamblin:1999by}, obtained by simply adding an extra warped direction to the 4-dimensional Schwarzschild.
\begin{figure}
\begin{center}
\includegraphics[width=1.7in]{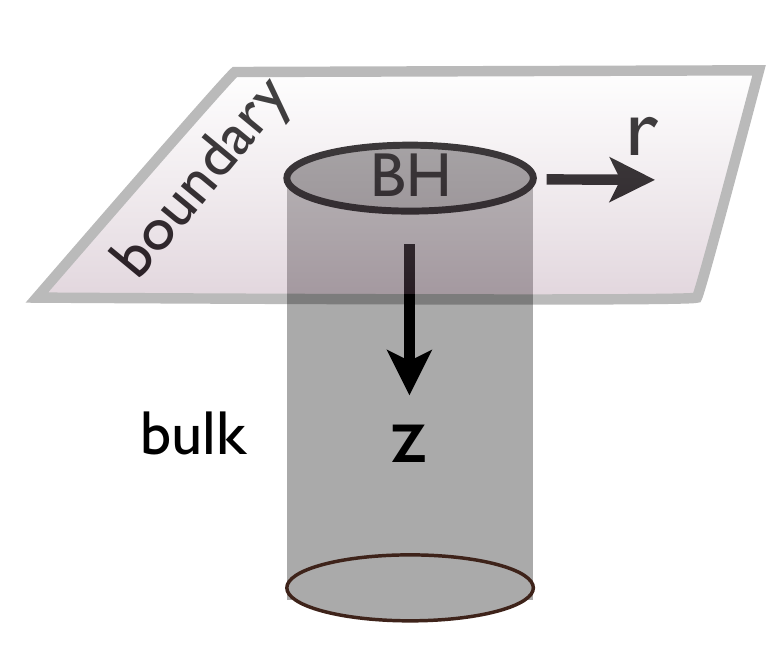}
\caption{Sketch of the AdS black string (the metric is given in \req{BSmet}).  The boundary radial distance $r$ is plotted horizontally, while the bulk radial coordinate $z$ is plotted vertically, with $z=0$ corresponding to the boundary.  At each constant $z$ slice, the induced metric is conformal to a 4-dimensional Schwarzschild black hole with horizon radius $r_+$.  Since the boundary metric is asymptotically flat, the bulk metric has a Poincar\'e horizon, which occurs at $z \to \infty$.
}
\label{f:BlackStringAdS}
\end{center}
\end{figure}
The geometry is sketched in \fig{f:BlackStringAdS}, where the boundary black hole (characterized by the requisite variation in the boundary radial coordinate $r$) extends `trivially' into the bulk (the bulk radial direction being labeled by $z$ with $z=0$ corresponding to the boundary).
However, while this is an exact solution to 5-dimensional Einstein's equations with negative cosmological constant, this solution suffers from a number of drawbacks.  
Most conspicuously, it is singular on the Poincar\'e horizon, since the cross-sectional area of the horizon scales as $1/z^2$ and therefore vanishes at $z \to \infty$.  Moreover, since sufficiently deep in the bulk the string becomes much thinner than the AdS scale, it suffers from a Gregory-Laflamme type instability \cite{Gregory:2000gf}, dynamically driving the string towards breaking up.
Most pertinently, it is easy to check that within the corresponding field theory state, the stress tensor induced on the boundary by the black string bulk geometry is not asymptotically thermal: classically it vanishes, and in terms of the field theory degrees of freedom, it does not grow with $N$ but rather  remains of $\ord{1}$.  As a result, these solutions appear to represent a confined phase of the dual field theory in the black hole background.\footnote{
The previous discussion also has a brane-world counterpart; see e.g.\ \cite{Fitzpatrick:2006cd}.  
}
In contrast,  we wish to study Hartle-Hawking-like states of the field theory on black hole spacetimes, such that the induced stress tensor scales as $\ord{N^2}$, corresponding to the deconfined phase. 

To guide our intuition, let us first consider what type of geometry would correspond to the deconfined phase if the field theory lived simply on the flat Minkowski spacetime. 
In fact, since the only scale in the problem is the temperature, there cannot be a phase transition at non-zero temperature, and the field theory always prefers to live in the deconfined phase. Formally, however, one can still consider the two distinct phases and their dual geometries.  These are
\begin{itemize}
\item 
pure Poincar\'e AdS$_5$ spacetime, which is dual to the (unstable) confined phase:
\begin{equation}
ds^2_{\text{planar AdS}} = \frac{1}{z^2}\, \left(-dt^2 + dx_i \, dx^i  + dz^2  \right)  \ ,
\label{planarAdS}
\end{equation}	
\item 
the planar \SAdS{5} black hole, dual to the deconfined phase:
\begin{equation}
ds^2_\text{planar BH} = \frac{1}{z^2}\, \left(- \left( 1 -\frac{z^{4}}{z_+^{4}} \right)\, dt^2 + dx_i \, dx^i   + \frac{dz^2}{1 -\frac{z^{4}}{z_+^{4}}} \right) \ .
\label{planarBH}
\end{equation}	
\end{itemize} 
Here  $t,x^i$ coordinates extend along the boundary at $z=0$.  The black hole horizon in \req{planarBH} is located at $z=z_+$ and has temperature $T = \frac{1}{\pi\, z_+}$.  The corresponding stress tensor of the large $N$ field theory is\footnote{
This boundary stress tensor (aka ``holographic stress tensor'') of the bulk solution \cite{Henningson:1998gx,Balasubramanian:1999re} is readily computed (following, say, \cite{deHaro:2000xn}) from the Fefferman-Graham expansion \cite{FG}; i.e., by expanding the metric in powers of $z$ in the above coordinates.}
 (see for example \cite{Bhattacharyya:2008jc}):
\begin{equation}
T^\text{planar  AdS}_{\mu\nu}  = 0 ,  \qquad
T^\text{planar  BH}_{\mu\nu} = \frac{(\pi \, T)^4}{16 \pi G_N^{(5)}}\, \left(\eta^{\mu\nu} + 4 \, \delta^\mu_t \, \delta^\nu_t\right).
\end{equation}

\paragraph{Expectations for bulk geometry:}
Returning now to the primary case of interest, wherein the field theory lives on an asymptotically flat black hole background, we expect that far from the black hole, the field theory stress tensor for the deconfined phase should behave as it would in flat spacetime.  In particular, the Euclidean path integral tells us that $T_{\mu \nu}(r)$ should approach the stress tensor of a thermal fluid in the asymptotic region $r \to \infty$.  As a result, the AdS$_5$ bulk dual should approach the planar AdS black hole (\ref{planarBH}) at large $x^i$, with a corresponding horizon localized near $z = \frac{1}{\pi \,T}$.   Furthermore, in any  region near the AdS$_5$ boundary ($z=0$), there must be a horizon, whose restriction to the AdS boundary coincides with that of the boundary black hole.  In other words, the event horizon of the boundary metric must extend into the bulk.  

To proceed further, let us consider how one can most naturally interpolate between an AdS black string near the AdS boundary (small $z$) and a planar AdS black hole in the boundary-asymptotic region (large $r$).
The most natural guess, inspired by using the fluid/gravity correspondence, would indicate that the `height' of the horizon $z$, as a function of boundary radius $r$, is determined by the local temperature $T_{loc}$ of the boundary stress tensor.
If we assume the boundary stress tensor corresponds to a static thermal fluid at temperature given by the local temperature of the Schwarzschild black hole with Hawking temperature $T$, this  is given by
\begin{equation}
T_{loc}(r) = {T \over \sqrt{\abs{g_{tt}(r)}}} 
=\( 4 \pi \, \rh \sqrt{1 - {\rh \over r}} \)^{-1} \ .
\label{Tlocal}
\end{equation}	
As long as the fluid/gravity intuition remains valid, this would suggest that the `height' of the bulk horizon $\zh(r)$ is given by 
\begin{equation}
\zh(r) \approx \frac{1}{\pi T_{loc}(r)} = 4 \,\rh \, \sqrt{1 - {\rh \over r}} 
\label{horshape}
\end{equation}	
%
\begin{figure}
\begin{center}
\includegraphics[width=3in]{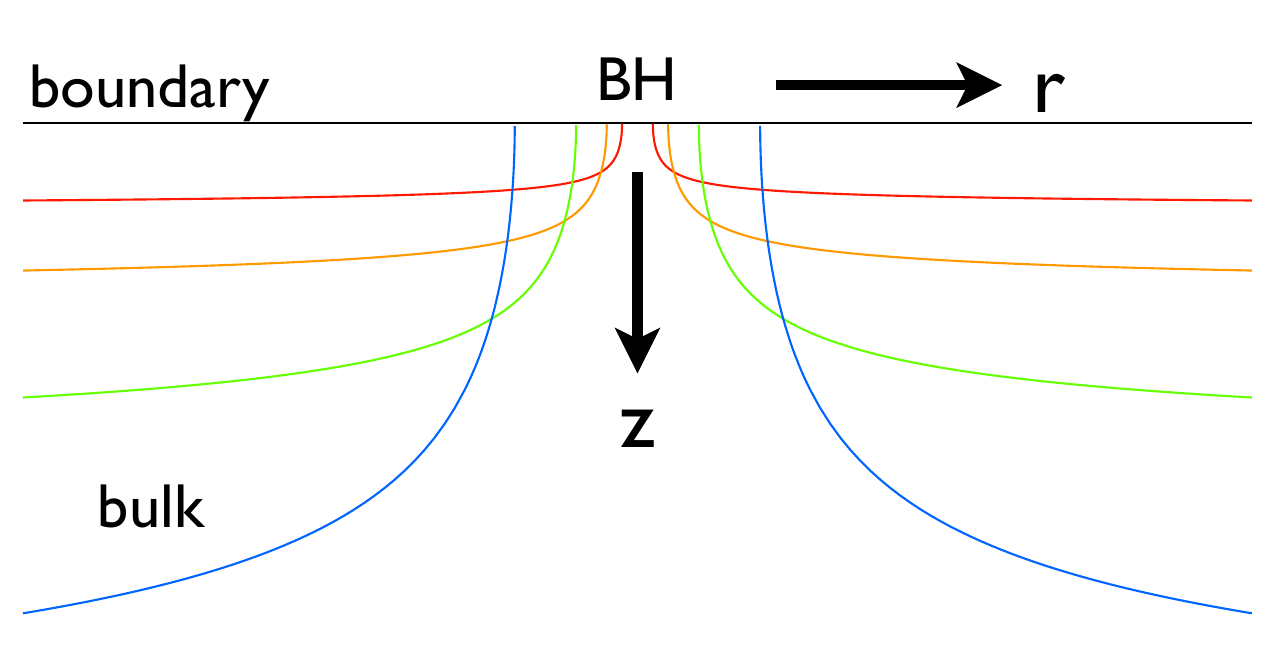}
\caption{Plot of the naive shape of the horizon of the Black Funnel, determined by the local temperature.  The AdS boundary is drawn as the black horizontal line, and the bulk extends below this line.}
\label{f:FunnelGuess}
\end{center}
\end{figure}
The full function is plotted in \fig{f:FunnelGuess} for several different values of $\rh$.
In \cite{Hubeny:2009ru} and subsequently \cite{Hubeny:2009kz,Hubeny:2009rc}, we dubbed a solution with this type of horizon shape a {\it black funnel}.
By construction, this spacetime has a correct asymptotic behaviour as $r \to \infty$, and it has a connected event horizon.
Note that since there is only one scale in the problem (the size of the boundary black hole, $\rh$), all the horizon shape functions are just scaled versions of each other -- as is evident from \req{horshape}.  In particular, the larger the boundary black hole $\rh$, the lower its temperature, so the further into the bulk the `shoulders' of black funnel reach.  

Before discussing the physical implications of such a spacetime, let us pause to consider whether this naive guess was the only sensible one.  In particular, are there other natural possibilities for what a horizon shape could be, subject to the boundary conditions specified above.  Some qualitatively different shape from the funnel (e.g.\ one with significantly richer structure or more prominent features) seems rather unnatural if the horizon remains connected.  However, we should question this assumption in turn.  Specifically, could there be two disconnected horizons?  We now argue that this is indeed likely to happen under certain conditions.  

\begin{figure}
\begin{center}
\includegraphics[width=4in]{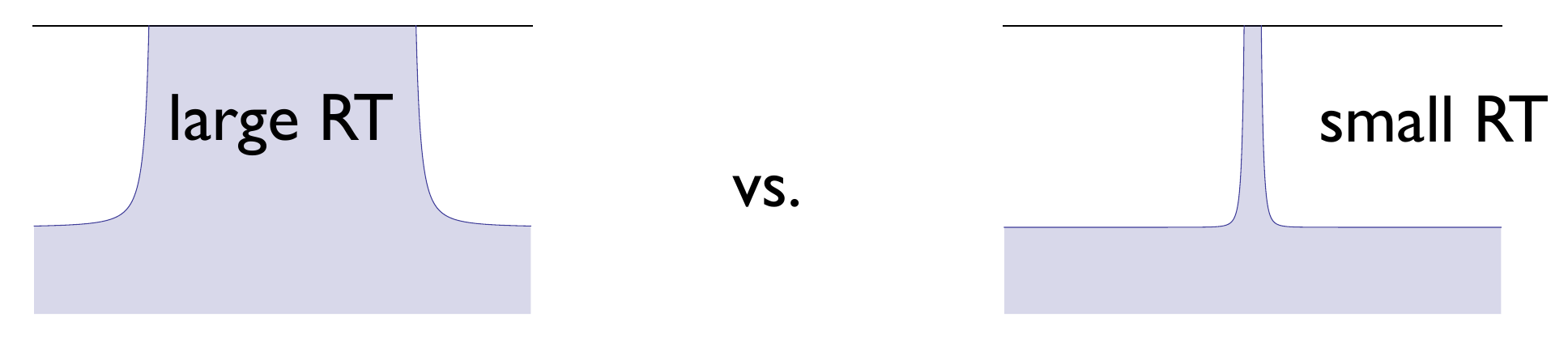}
\caption{Illustration to indicate why a funnel configuration is most natural for large $R \, T$:  for small $R \, T$ the funnel might be very thin and elongated, and therefore likely unstable to pinching off.}
\label{f:LargeSmallRT}
\end{center}
\end{figure}
Consider fixing the temperature $T$ of the boundary black hole, but vary its size, which we will henceforth denote as $R$ to distinguish it from the horizon radius in Schwarzschild background.  When $R \, T$ is large, we would expect to obtain a stubbier funnel, such as sketched in the left panel of \fig{f:LargeSmallRT}.  Such a solution seems perfectly acceptable and stable.  On the other hand, when $R \, T$ is small, a funnel-type solution would force the funnel's throat to be highly elongated, as indicated in the right panel of \fig{f:LargeSmallRT}.    This looks unstable to the horizon pinching off, analogously to the Gregory-Laflamme instability \cite{Gregory:1993vy,Gregory:2000gf} mentioned above.
If the horizon actually pinches off, there would be two\footnote{
One could also imagine more components of the horizon, originating from multiple pinching.  However, it is unlikely that such a solution would be static -- the disconnected horizon `drops' would presumably fall down onto the `bottom' horizon, probably settling down to the two-horizon solution.
}
disconnected horizons:  the one ending on the boundary black hole, and the one which extends to large $x_i$ and gives rise to the thermal stress tensor asymptotically.  Indeed, one can postulate such a solution by approaching the argument from the other direction.
Start with  a planar black hole (\ref{planarBH}) at some temperature $T = T_H$, which is stable to small perturbations, and perturb this solution by adding a black hole of the same temperature $T_H$ to the boundary metric.  Our general experience with AdS spaces suggests that the effect of a disturbance of size $R$ on the boundary falls off deep in the bulk and typically extends to $z \sim R$.  
 Thus, since the planar black hole horizon is at $z = \frac{1}{\pi\, T_H}$, the effect on this horizon of adding the black hole to the boundary should be perturbative for $R \, T_H\ll \frac{1}{\pi}$.  In this case we expect to find a static bulk solution with two disconnected horizons: one horizon will be a perturbed planar black hole, while the other should hang down into the bulk from the boundary black hole, with the horizon capping off smoothly far above the (perturbed) planar horizon.    We refer to the first horizon as simply the planar black hole, while the second horizon we christen a {\it black droplet.}  On the other hand, if we increase $R \, T$ enough for the two horizons to touch, we would expect them to merge into a funnel-type solution.

In \fig{f:FunnelDroplet} we sketch the two types of solution discussed so far: black funnel on the left and black droplet above a deformed planar black hole on the right.  Both of these solutions should be relevant in their corresponding regime.  In particular, the preceding arguments motivate the following conjecture \cite{Hubeny:2009ru}:
\begin{figure}
\begin{center}
\includegraphics[width=5in]{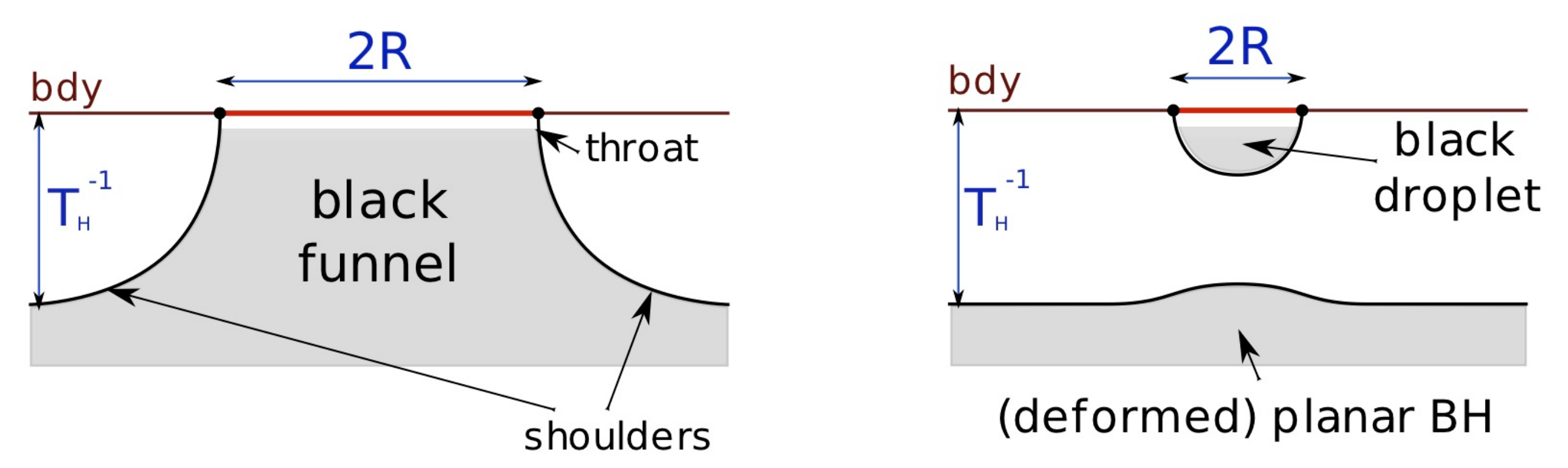}
\caption{A sketch of our two novel classes of solutions dual to a deconfined phase of a strongly coupled field theory on a black hole background: {\bf (a):} black funnel and  {\bf (b):} black droplet above a deformed planar black hole. The shaded part indicates regions which are behind horizons in the bulk spacetime, and the thick (red) line indicates the black hole of the boundary metric.}
\label{f:FunnelDroplet}
\end{center}
\end{figure}

\paragraph{Conjecture:}
Deconfined phase in a strongly coupled CFT on a black hole background has
the corresponding bulk dual geometry described by a
black funnel, or a black droplet suspended over
a deformed planar black hole (depending on $R \, T$).


What would it take to confirm our conjecture?
Ideally, we would like to obtain the requisite solutions in analytical form.
We can write the general form of the bulk metric, e.g.\ in Fefferman-Graham coordinates, as 
\begin{equation}
ds^2 = {1 \over z^2} \, \left[ - f(r,z) \, dt^2 + g(r,z) \, dr^2 
+ h(r,z) \, r^2 \, d\Om_2^2  + dz^2 \right]
\label{genmet}
\end{equation}	
where we have explicitly set the AdS radius to unity, which corresponds to measuring all distances in AdS units.  It is straightforward to write Einstein's equations for this ansatz -- but far too hard to solve them: we need to find the functions $f$, $g$, and $h$, satisfying the following boundary conditions on the metric form:
\begin{itemize}
\item 
The metric must approach that of the Schwarzschild black string near the boundary, i.e.\ as $z \to 0$,
\begin{equation}
ds^2 \to   {1 \over z^2} \, \left[ - \left( 1-{\rh \over r} \right) \, dt^2
 + {dr^2 \over  1-{\rh \over r}} + r^2 \, d\Om_2^2  + dz^2 \right]
\label{BSmet}
\end{equation}	
\item
Asymptotically in the boundary-radial directions, i.e.\ as $r \to \infty$, the metric must approach that of the black brane in AdS \req{planarBH}, which by coordinate-transforming 
$z \to z/\sqrt{1+{z^4 \over \zh^4}}$ to render it in the form \req{genmet}, we can express as
\begin{equation}
ds^2 \to {1 \over z^2} \, \left[ - {\left(1-{z^4 \over \zh^4} \right)^2 \over \left(1+{z^4 \over \zh^4}\right)} \, dt^2 + \left(1+{z^4 \over \zh^4} \right) \, \left(dr^2 
+ r^2 \, d\Om_2^2 \right) + dz^2 \right]
\label{BBmet}
\end{equation}	
\end{itemize}
The parameters $\rh$ and $\zh$ are related by requiring that the temperature is constant on the entire horizon, 
\begin{equation}
T = \frac{1}{4 \, \pi \, \rh} = \frac{\sqrt{2}}{\pi \, \zh} \ . 
\label{}
\end{equation}	
Although such a solution is much too difficult to find analytically, it should be possible to find numerically; in fact work is underway to construct such solutions.

Even though direct verification of our conjecture by finding funnel and droplet solutions analytically is presently out of reach, we nevertheless have strong evidence for the conjecture from analogous solutions in  related context.
One set of examples arises from lower-dimensional settings \cite{Hubeny:2009ru}, where finding exact solutions is easier.   A rather trivial example to consider is an asymptotically flat 1+1 dimensional boundary black hole.  Since there is no way to `go around' such a black hole, these black holes should be thought of as having a large size $R$ in comparison with any other scale.  As a result, only black funnel solutions would be expected to exist, which is indeed borne out by explicit calculation.  The requisite funnel solution is obtained by a suitable change of coordinates from pure AdS$_{3}$ (see \cite{Hubeny:2009ru} for details).

Turning to higher dimensional examples, \cite{Hubeny:2009ru} also considers 2+1 dimensional black holes on the boundary with asymptotically locally AdS$_{4}$ bulk dual.  Here we were able to write explicit solutions by exploiting the known 3+1 dimensional AdS C-metric solutions, further discussed in \cite{Hubeny:2009kz}.   However, although in  this context we find both black funnels and black droplets suspended above disconnected horizons, they are not dual to field theory on asymptotically flat black hole: 
instead, we can only obtain 2+1 dimensional boundary black holes which live in spatially-compact geometry or are asymptotic to $H^2 \times {\mathbb R}$  (where $H^2$ is the hyperbolic plane) within the AdS C-metric family of solutions. 
Moreover, within this family, the two disconnected horizons in the droplet case cannot attain the same temperature, so that they cannot correspond to the thermally-equilibrated Hartle-Hawking states.\footnote{
As was recently shown \cite{Caldarelli:2011wa} in the context of charged AdS C-metrics, adding an extra charge parameter enables the horizons in the droplet configurations to attain same temperature on both horizons, but not simultaneously the same chemical potential. 
}
Pursuing further the exploration of boundary metric with negative curvature, \cite{Hubeny:2009rc} uses a trick of double-Wick rotation to describe analogs of funnels describing a Hartle-Hawking state on \SAdS{4} boundary metric.  However, here the nature of confined versus deconfined states is far more subtle, and will not be discussed further in this talk.

\paragraph{Qualitative predictions:}
Switching focus from evidence for our conjecture to its implications, let us assume that the conjectured funnels and droplets indeed exist, and examine what would be the consequences.
We have argued that both bulk configurations -- black funnel  and the black droplet suspended above a deformed planar black hole -- correspond to a deconfined phase in the CFT, having the requisite asymptotically thermal stress tensor (scaling like the central charge, $\CO(N^2)$, of the CFT).  However, depending on which geometry is realized, the physical properties, such as the response of the system to perturbations, differ significantly \cite{Hubeny:2009ru}.  
 From the field theory viewpoint, which possibility occurs depends on whether the field theory black hole couples strongly or weakly to the plasma excitation, in the following sense:
 
The CFT dual of a black funnel represents a black hole in equilibrium with a deconfined plasma to which it is strongly coupled, in direct large-$N$ analogue of the free Hartle-Hawking state.  To see this, consider such a state at finite volume (e.g., on a torus).  If the volume is either increased or decreased by a small amount, one expects heat to flow either out of or into the black hole so that the solution quickly settles down to a new Hartle-Hawking-like state at the same temperature.  In the bulk, this flow would propagate along the funnel horizon.
In contrast, the two-horizon solutions present a greater surprise.  They represent black holes in equilibrium with a deconfined plasma to which they are only very weakly coupled.  One notes that the plasma excitations, thought of as small deformations of the bulk planar black hole, can pass directly under the black droplet with minimal interaction. The two horizons can exchange heat only via the $\ord{1}$ fields that propagate in the bulk.  In the previous gedanken experiment, the bulk AdS solution clearly responds by increasing or decreasing the height of the planar black hole horizon; i.e., by heating or cooling the plasma so that its temperature $T$  no longer matches $T_H$ of the boundary black hole.  It is only over very long time scales that quantum effects in the bulk will return the plasma to the black hole temperature $T_H$.  This slow rate of heat exchange with the field theory black hole is more typical of what one would expect in a {\it confined} phase.

\begin{figure}
\begin{center}
\includegraphics[width=4in]{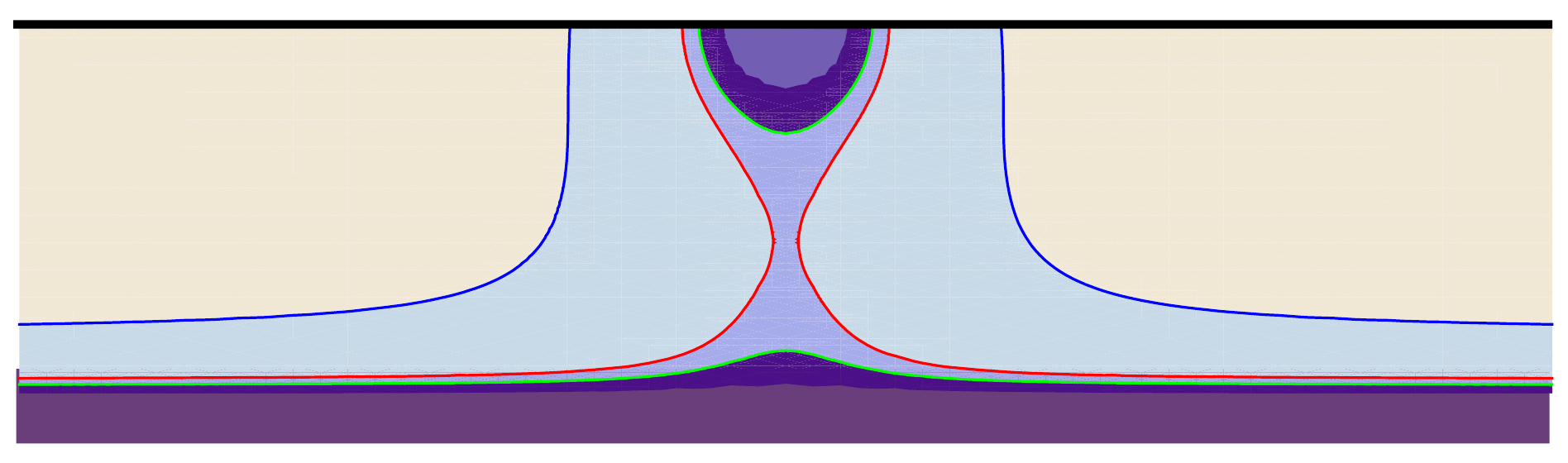}
\caption{A cartoon of possible transition between a funnel and a droplet.  The blue (outer) contour corresponds to the funnel configuration at large $R \, T$, the green (inner, disjoint) set of contours bordering the dark purple region corresponds to a droplet above a deformed planar black hole occurring for small $R \, T$, while the red contour corresponds to a natural configuration close to the transition point at $R \, T \sim 1$. }
\label{f:FunDropTrans}
\end{center}
\end{figure}

Consequently, as the parameters $R \, T$ of the background are varied, we expect a new kind of phase transition to occur in the large-$N$, strong coupling ($\lambda \gg 1$) limit.\footnote{
Due to the change in connectivity of the bulk horizon, the on-shell action in the bulk will be non-analytic in parameters that control the size of the boundary black hole. This then implies that the partition function of the field theory in the boundary black hole background will exhibit non-analytic behavior characteristic of phase transitions. 
}
It would be very interesting to probe the nature of this transition.  In particular, one might expect this transition to be similar to the Gregory-Laflamme transition, possibly to display corresponding singular solutions with ``pinched horizons"  connecting parts of the phase diagram with different topologies. This possibility is indicated in \fig{f:FunDropTrans}.
If the funnel does pinch off as $R \, T$ is decreased to some critical value, does this result in some interesting critical behaviour?  Furthermore, although we started out by motivating the funnel solution by considering CFT on Schwarzschild background, given that in this case $R \, T \sim 1$, we might well ask, which side of the phase transition does Schwarzschild actually lie on?  I.e.\ is the bulk dual a funnel or a droplet?

\paragraph{Puzzle:}
Let me close off this part of my talk with a puzzle, concerning stationary but non-static boundary black hole metrics.  Suppose we take one of our boundary black holes for which the bulk dual is described by a funnel geometry, and `spin it up' slightly.
From the field theory viewpoint, rotating asymptotically flat field theory black holes do not admit well-defined Hartle-Hawking states \cite{Kay:1988mu} due to features associated with super-radiance.  Nevertheless, in keeping the boundary geometry stationary, we would expect the state of the CFT, and consequently its bulk dual, to be stationary. 
However, from the bulk perspective, the dual geometry cannot be described by a rigidly-rotating funnel, for that would imply superluminal horizon velocity far enough down the funnel flanks.  If, on the other hand, the funnel does not rotate rigidly, then this would suggest the first example (which I am aware of) of a stationary black hole geometry with its event horizon {\it not} coinciding with a Killing horizon.\footnote{
While stationary black holes without Killing horizons are unfamiliar, there is no contradiction with the known rigidity theorems (e.g.\ \cite{Hollands:2006rj}) which require a compactly generated horizon.   Nevertheless, even so, it is worth stressing that this would be a very bizarre set-up from the GR perspective.
}

Another example of this remarkable occurrence should arise in black funnel solutions displaying a constant flow of heat into or out of the boundary black hole.  One could fix the temperature of the shoulder region of \fig{f:FunnelDroplet} independently of the boundary black hole temperature $T_H$.  This is again a stationary situation, in which one expects a stationary bulk AdS solution.  However, here too, the event horizon cannot be a smooth Killing horizon, as such horizons have uniform temperature.   
The possibility of such novel solutions raises many further questions about the relation between the boundary heat flow and the bulk horizon generators, compatibility with Raychaudhuri's theorem, and the bulk description of the flows of negative energy across the horizon of the field theory black holes that one associates with Hawking radiation in the field theory.  

\section{Field theory with time-dependence}
\label{s:funnels}

So far, we have seen above that even innocuous static configurations can lie outside the `long-wavelength' regime of fluid/gravity correspondence, by virtue of living on a background which is too curved.  Let us now switch gears and consider configurations that, while on flat background, exhibit sufficient time-dependence to cast them outside the long-wavelength regime.  In general, a typical dynamical process involving a black hole (such as its formation or coalescence, quasinormal ringing, etc.) occurs on time scales set by the black hole size, i.e.\ the thermal scale.  Nevertheless, there are interesting and important physical processes which may be at least partially understood from the fluid/gravity framework.  
In the following discussion, we will only consider systems with `mock' time-dependence given by simply taking a static configuration in different coordinates.  The presentation is based primarily on \cite{Figueras:2009iu}, which in addition discusses several configurations which are genuinely time-evolving.  We will see that even the `secretly-static' class offers intriguing surprises.

\paragraph{Conformal soliton:}
One of the simplest examples of a secretly static configuration which nevertheless provides an instructive toy model for a time-dependent scenario is given by the {\it conformal soliton} geometry originally introduced in \cite{Friess:2006kw}, and then studied extensively in \cite{Figueras:2009iu}.   
The spacetime in question is simply a patch of global \SAdS{} black hole; so
its metric is known explicitly, and admits the corresponding
Killing fields.  Nevertheless, if we work in a coordinate system
such that the field theory lives on flat spacetime $\RR^{3,1}$ rather than $S^3\times \RR^1$, which means considering only a `Poincar\'e patch' of the \SAdS{} black
hole, the time translation symmetry is no longer manifest, and the
solution looks highly dynamical. 
\begin{figure}
\begin{center}
\includegraphics[width=1.3in]{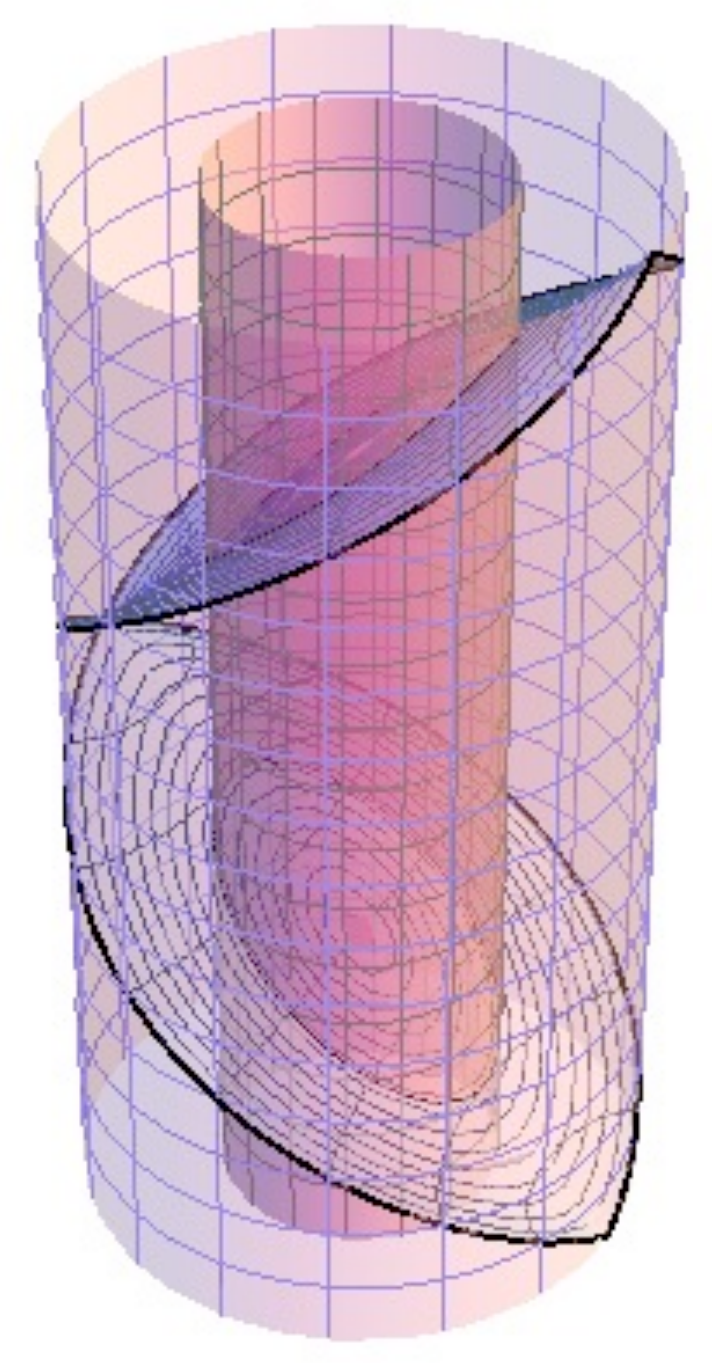}
\hspace{1.5cm}
\includegraphics[width=3in]{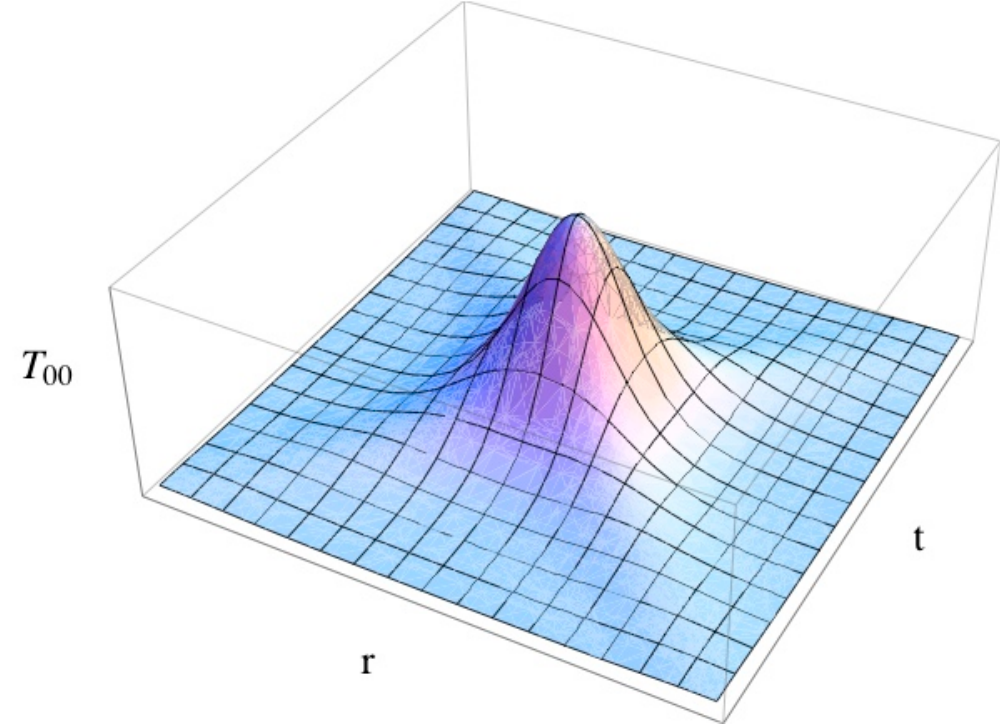}
\caption{Conformal Soliton geometry  (left) and dual field theory energy density (right). 
On left, Poincar\'e wedge of Schwarzschild-AdS geometry is indicated.  The inner cylinder is the global event horizon as in \fig{f:HawkPage}, the outer cylinder the AdS boundary, and the diagonal surfaces indicate the Poincar\'e AdS horizon $t=\pm \infty$.
On right, $T_{00}(t,r)$ is plotted as a function of boundary time $t$ and boundary radial variable $r$.
}
\label{f:ConfSolGeom}
\end{center}
\end{figure}
Pictorially, it describes a
black hole entering through the past Poincar\'e horizon and exiting
through the future Poincar\'e horizon, with its closest approach to
the boundary occurring at $t=0$ (Poincar\'e time).  
This is indicated in the left panel of \fig{f:ConfSolGeom}, where the Poincar\'e patch is shown on global Schwarzschild-AdS.

In the boundary CFT, this configuration corresponds to
a finite energy lump which collapses and re-expands in a
time-symmetric fashion.  For illustration, the energy density is plotted in the right panel of \fig{f:ConfSolGeom}. 
An important point of note is that although the fluid dynamical approximation does not remain
valid for all times, because this fluid flow is conformal to a
stationary fluid on the Einstein static universe, the stress tensor is
shear-free; that is, there is no dissipation in this fluid flow.
Another way to say this is to observe that the static Schwarzschild-AdS black hole describes a static ideal fluid in the boundary CFT on $S^{3}\times \RR^1$.  Since ideal fluids admit no dissipation, there is no entropy production in such a system.  Restricting this system to only the Poincar\'e patch corresponds to performing a conformal transformation of the boundary metric to $\RR^{3,1}$.  However, since entropy is invariant under a conformal transformation, it remains constant, confirming our statement that there is no entropy production.  The total entropy for the conformal soliton flow is then given by quarter of the global \SAdS{} event horizon area in Planck units.

\begin{figure}
\begin{center}
\includegraphics[width=1.5in]{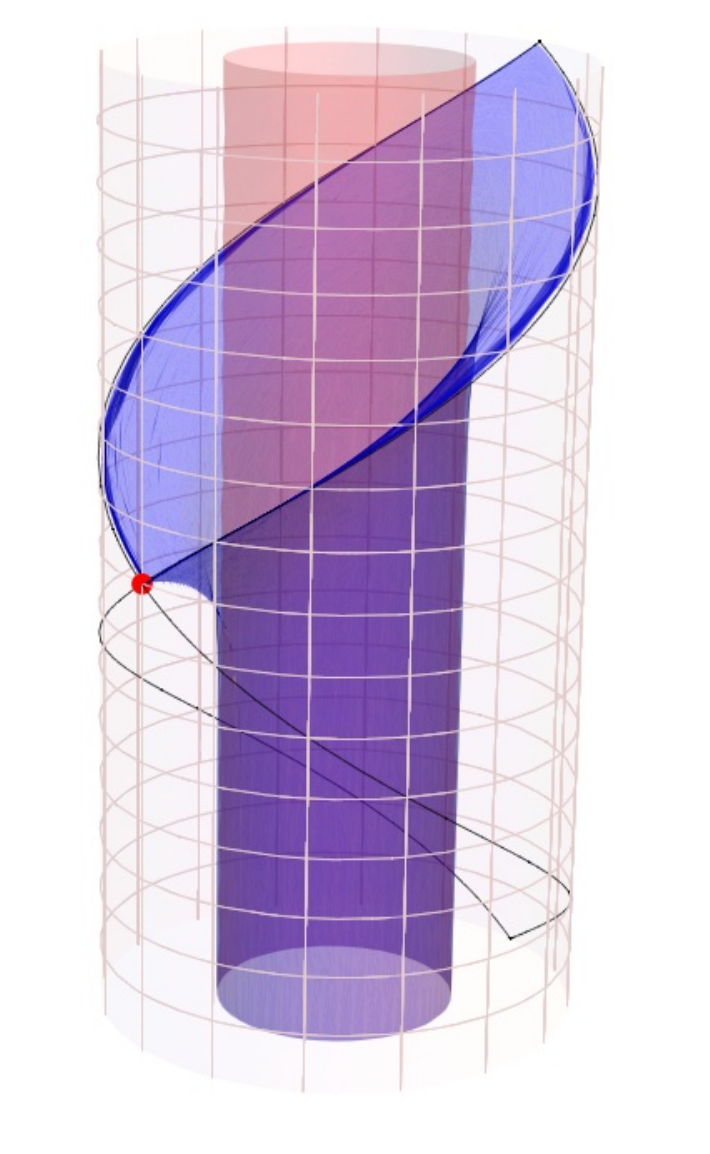}
\caption{The conformal soliton horizons: the outer cylinder represents the AdS boundary, the inner (red) cylinder corresponds to the global \SAdS{} event horizon, which as we argue below coincides with the  conformal soliton's apparent horizon, whereas the flared (blue) surface is the actual event horizon.}
\label{f:CShorizon}
\end{center}
\end{figure}

\paragraph{Event horizon:}
Let us now consider the event horizon of the conformal soliton geometry.
Naively one might
expect that since we are just performing a coordinate transformation
on a given solution, any geometrical feature, such as the location of
the event horizon, remains invariant under such a transformation.  This would lead us to expect that the event horizon of the Poincar\'e
patch of \SAdS{} black hole coincides with the event horizon of the
global \SAdS{}.  However,  this is not the case!
Recall that the event horizon $\CH^+$ is defined as the boundary of the past of the future null infinity $\scri^+$.
Since we are restricting consideration to a subregion $\scri^+_{CS}$ of the full
future null infinity $\scri^+$ of the global \SAdS{} spacetime, the actual event horizon
for the conformal soliton lies outside the global event horizon of
\SAdS{}. 
It is in fact easy to calculate the location of the event horizon explicitly (which is achieved simply by tracing certain null geodesics in our geometry, as explained in \cite{Figueras:2009iu}); doing so,
we find that it starts out near the
global \SAdS{} event horizon at very early times and `flares out' towards the AdS Poincar\'e horizon close to the boundary.
This is plotted in \fig{f:CShorizon}, which shows the event horizons for both global \SAdS{} and the conformal soliton.

Having found the location of the event horizon, it is straightforward to find its area at any given Poincar\'e time slice.  
In fact, as is clear from \fig{f:CShorizon}, and was explicitly calculated in \cite{Figueras:2009iu}, the area of this conformal soliton event horizon diverges at late times.
This surprising result leads to a puzzle: if we associate the entropy
of the corresponding CFT conformal soliton state to the area of the conformal soliton
event horizon, as is usually assumed to be the case, then we find that
this entropy likewise diverges at late times.  But as we have already argued, the CFT entropy is finite and constant in time. 
This implies that {\it the event horizon area does not correctly reproduce the CFT entropy!}

\paragraph{Dual of CFT entropy?}
This observation begs the obvious question: is there a geometric quantity in the bulk which does reproduce the CFT entropy of the corresponding state?
We are used to thinking of the black hole's horizon area as capturing the number of its microstates and hence the bulk entropy, to be identified with the boundary entropy.  However, in retrospect, associating the event horizon area to entropy is a very strange thing to do: the event horizon, as indicated above, is defined in a highly teleological manner: its location at a given time depends on the entire future evolution of the spacetime.  On the other hand, we think of the entropy as a temporally-local attribute, describing the state independently of what happens to it in the future.  This suggests that perhaps we should turn to more local concept of the black hole, one described by a quasi-locally defined horizon, as a possible candidate for capturing the CFT entropy.
The most natural guess would then be the apparent horizon.\footnote{
In fact, this idea is not new:
Already prior to \cite{Figueras:2009iu}, it has been argued in several different contexts
(e.g.\ \cite{Hubeny:2007xt,Chesler:2008hg}) that it may be more appropriate to
associate the entropy of the CFT configuration to the area of the
apparent horizon, rather than the event horizon, in the bulk dual.}

\paragraph{Apparent horizon:}
To test this possibility, we now turn to considering the location of an apparent horizon in the conformal soliton geometry.  Since the notion of an apparent horizon is intimately connected to that of trapped surfaces, let us first recall the definition of the latter.  Given a foliation of the spacetime, a {\it trapped surface} $\CS$ is a closed surface (within a given leaf of the foliation) with negative divergence $\theta$ for both outgoing and ingoing null congruence normal to $\CS$, i.e.\ the areas of the `wavefronts' for both of these null
congruences decrease in time.  Physically, the presence of such a
trapped surface indicates a region of strong gravitational effects; indeed, in physically reasonable spacetimes such as the presently considered one, any trapped surface must be contained within a black hole event horizon.
A surface is marginally trapped if the outgoing null congruence has
zero expansion, while the ingoing congruence has negative expansion.
The {\it apparent horizon} is then defined as the boundary of the trapped surfaces, or outermost marginally trapped surface.\footnote{
These two notions are strictly-speaking different, but coincide in a wide class of physically reasonable spacetimes, including the present one.
For a good review of various notions of quasi-local horizons, see \cite{Booth:2005qc,Ashtekar:2004cn}.  Technically speaking,  an apparent horizon is defined as a co-dimension 2 surface, on a given leaf of foliation; however, here we will stick to the more commonly used terminology in non-GR literature of apparent horizon signifying the full co-dimension 1 tube, obtained by putting together the outermost marginally trapped surfaces from all the leaves of the given foliation.  Technically speaking, this tube is called isolated horizon if it is null, and dynamical horizon if spacelike.
}

An important subtlety to note about both of these definitions is that
a given spacetime geometry does not by itself specify the location of
the apparent horizon; we first need to specify a foliation of the
spacetime, with respect to which we can then define the apparent
horizon.\footnote{
In fact, as demonstrated in \cite{Wald:1991zz}, even the
  Schwarzschild black hole spacetime admits (sufficiently bizarre)
  foliations for which there are no trapped surfaces at all, so that
  there is no apparent horizon.}  In the present case, the most natural foliation would be the one corresponding to constant Poincar\'e time
slices, but this statement is nevertheless unappealingly coordinate-dependent.   
For general spacetimes, changing the foliation slightly could correspondingly change the location of the apparent horizon slightly.  However, as argued in \cite{Figueras:2009iu},  this is not the case for stationary black holes, where the event horizon is a Killing horizon.
In particular, one can show that if the spacetime admits a Killing horizon, then this Killing horizon is an apparent horizon for any foliation which contains full slices of the horizon.\footnote{
The gist of the argument is the following:
Any slice of a Killing horizon is a marginally trapped surface, since the outgoing null
normals to any such slice of the horizon coincide with the horizon
generators (due to the Killing horizon being null), and the horizon
generators have zero expansion (because any spacelike slice of a
Killing horizon has the same proper area).  Moreover, this marginally
trapped surface is the outermost one, since there cannot be trapped
surfaces outside the event horizon.  Hence for a slicing which admits
a complete cross-section of the (future) event horizon, the apparent
horizon necessarily coincides with the event horizon.  
}
This is the case for our conformal soliton geometry.

Hence the apparent horizon of the conformal soliton, e.g.\ in Poincar\'e slicing,
coincides with the global \SAdS{} event horizon, whose area is indeed
constant and given by the expected value.  Since the event horizon and the
apparent horizon behave radically differently (cf.\ \fig{f:CShorizon}), this conformal soliton geometry provides a
good testing ground for studying the distinction between the two horizons and the role they play for the associated CFT dual.
We see that in this case the CFT entropy is clearly more naturally associated with the
apparent horizon rather than the event horizon.

The large disparity between locations of the event and apparent horizons might seem enigmatic in light of \cite{Bhattacharyya:2008xc}, where it was argued that within the hydrodynamic regime, the event horizon and the apparent
horizon track each other closely.\footnote{
Further more recent discussions include \cite{Booth:2010kr,Booth:2011qy}.
}
The essential difference between these two contexts is that in \cite{Bhattacharyya:2008xc}, it was assumed that the geometry would settle down at late
times to a stationary finite-temperature black hole.   This is not the case for the conformal soliton geometry. It is precisely these late time boundary conditions that force the apparent
horizon and the event horizon to behave very differently.

\paragraph{Puzzle:}
We have seen that in certain cases of large disparity between apparent and event horizon, such as in our conformal soliton setting, the CFT dual entropy seems to be given by area of the apparent horizon, rather than the event horizon as conventionally  expected.
Since this striking observation is rather surprising and, if true in general, would have far-reaching implications, let us end our discussion by probing this issue a bit deeper.  In particular, can it be true that the entropy is associated with the area of the apparent horizon?  As we pointed out above, the fact that CFT entropy is not determined by the area of the event horizon in general is perhaps not so surprising, given that the event horizon is a global construct, while the CFT entropy is not.
On the other hand, there are nevertheless serious problems with associating the area of an apparent horizon, or any other quasi-local horizon, to the dual CFT state entropy.  One difficulty is that  the entropy (when defined) is expected to be smoothly varying in time, whereas the area of a given apparent horizon could in principle jump discontinuously. 
A more problematic issue, however, is that the location of apparent horizons in non-stationary spacetimes depends on a choice of spacetime foliation, whereas the CFT state at a given time carries no such information.
 This seems to imply that our bulk
prescription has more freedom or ambiguity in defining the entropy
than that afforded by the boundary theory. 

One possible resolution is that there is a preferred foliation of the bulk, such as a zero-mean-curvature slicing, on which one is supposed to evaluate the area.  However, we have no compelling justification for this option. A simpler resolution to this puzzle would be that in the regime where the concept of entropy is
meaningful, the horizon has to be evolving slowly enough that there is
negligible difference between the areas of all slices of horizon which
end on the same boundary time-slice.  This is essentially the same
picture as that advocated in \cite{Bhattacharyya:2008jc,Bhattacharyya:2008xc}, except that here we
use it for apparent horizon rather than the event horizon.
In effect the field theory on the boundary should achieve local equilibrium in order for entropy to constitute a meaningful observable.  (Note that there is no such restriction for e.g.\ the entanglement entropy associated with a given boundary region, whose proposed the bulk dual  \cite{Hubeny:2007xt} is related to the area of an extremal surface anchored on that boundary region.)

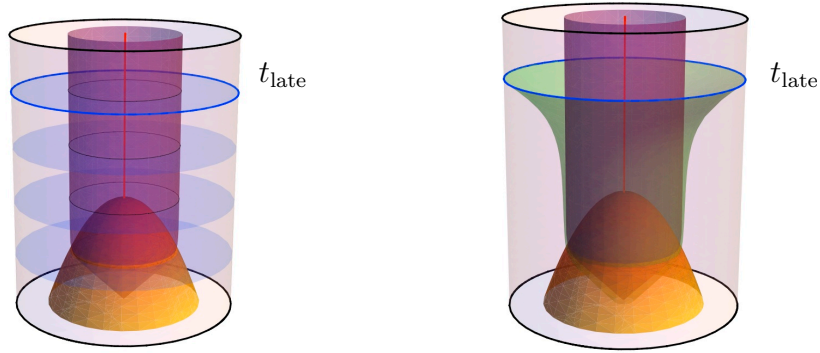
\begin{figure}
\begin{center}
\includegraphics[width=1.7in]{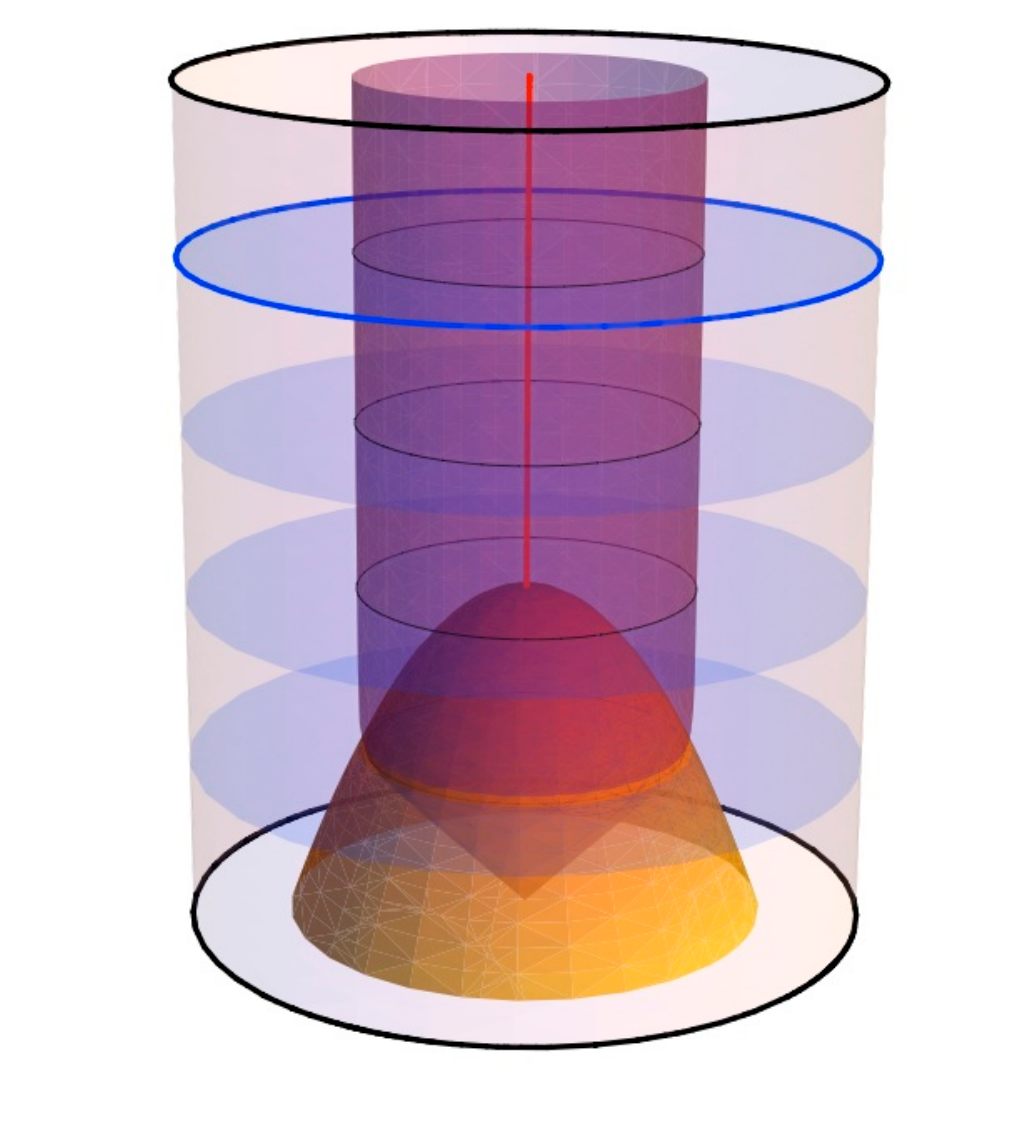}
\hspace{2cm}
\includegraphics[width=1.7in]{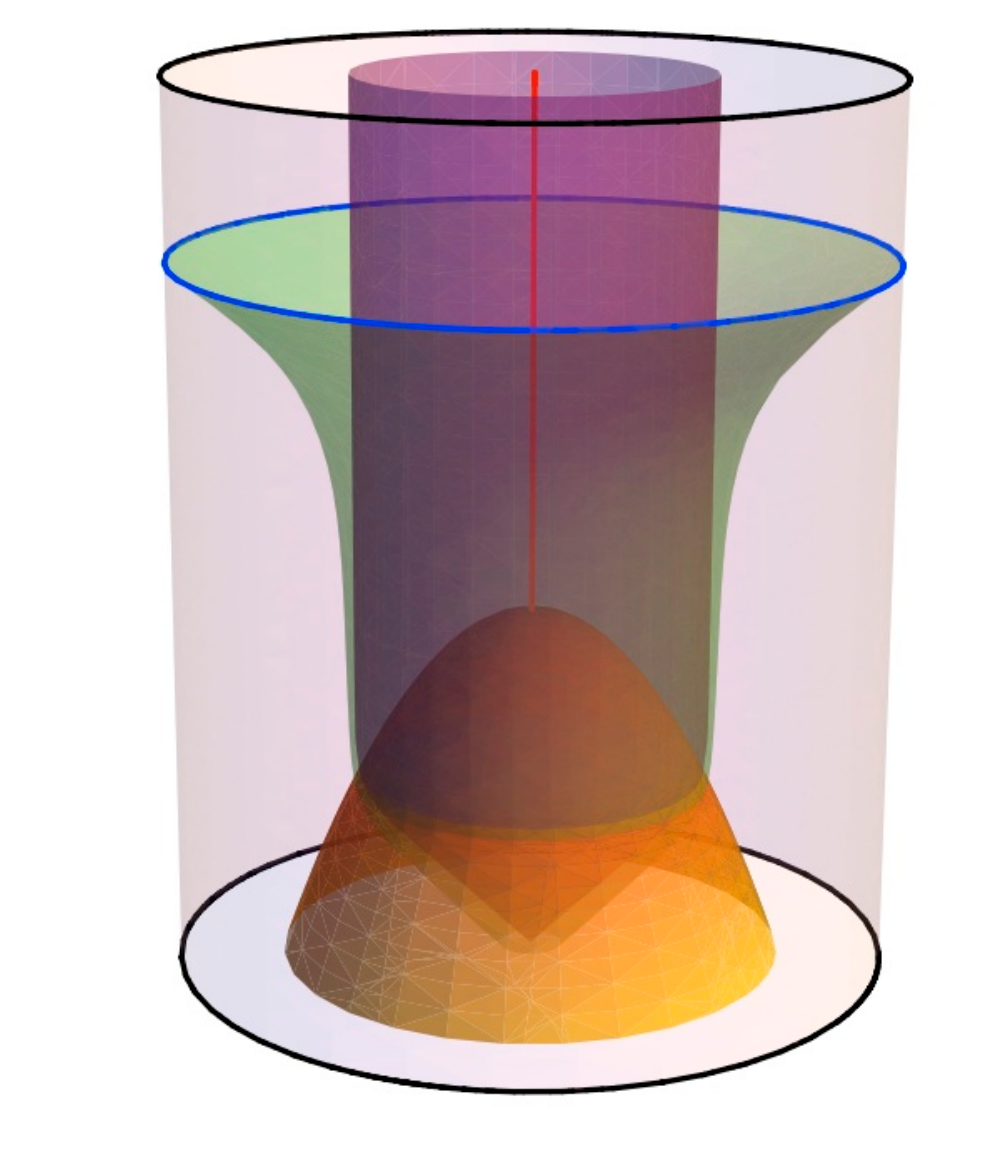}
\begin{picture}(0,0)
\setlength{\unitlength}{1cm}
\put(-7,3.9){$t_{\rm late}$}
\put(-0.2,3.9){$t_{\rm late}$}
\end{picture}
\caption{Possible spacelike foliations of a geometry a black hole collapse. The outer cylinder represents the AdS boundary; the bottom (orange) surface represents a collapsing star, the (red) vertical line the curvature singularity, the inner (purple) cylinder the event horizon.  Two distinct sets of foliations are sketched.  On {\bf left}, the corresponding apparent horizon has settled down at late time $t_{\rm late}$ (indicated in blue). On {\bf right}, the spacelike surface (green) at $t_{\rm late}$ admits no apparent horizon at all.}
\label{f:foliations}
\end{center}
\end{figure}
However, even this proposition cannot be the full story, as illustrated by the following scenario:
Suppose we collapse a black hole, and wait for very long time after the black hole has settled down, say time $t_{\rm late}$ indicated in \fig{f:foliations}.   A `natural' foliation (sketched in the left panel) would give the expected results: by the time $t_{\rm late}$, the apparent horizon area has grown and settled down to a constant  which coincides with the event horizon area.  However, we can devise a foliation (sketched in the right panel) which, even at $t_{\rm late}$, admits no trapped surfaces and therefore has zero apparent horizon area.  The corresponding spacelike surface `dips just below' the event horizon formation; and since the event horizon itself is null, the requisite foliation surface can always be constructed so as to be spacelike.
It would be interesting to understand the full implications of this observation, since it bears on the relation between boundary time and bulk time: which bulk region is most naturally associated with a given boundary time, such as $t_{\rm late}$?   

\subsection*{Acknowledgements}
\label{acks}

It is a pleasure to thank my collaborators,
Pau Figueras,
Don Marolf,
Mukund Rangamani,
and Simon Ross
for wonderful collaborations.
This work is supported in part by a STFC Rolling Grant.


\providecommand{\href}[2]{#2}\begingroup\raggedright\endgroup

\end{document}